\documentclass[conference]{IEEEtran}
\ifCLASSINFOpdf
\else
\fi
\usepackage[acronym,toc,shortcuts]{glossaries}
\usepackage{multirow}
\usepackage{amssymb}
\usepackage{multirow}
\usepackage{array}
\usepackage{caption}
\usepackage{lipsum}
\usepackage{mathtools}
\usepackage{cuted}
\usepackage{breqn}
\usepackage{graphicx}
\usepackage[english]{babel}
\usepackage[autostyle]{csquotes}
\usepackage{amsmath}
\usepackage{verbatim}
\usepackage[graphicx]{realboxes}
\usepackage{breqn}
\usepackage{subfig}
\usepackage{euscript}
\usepackage{color}
\usepackage[lined,linesnumbered,noend,ruled]{algorithm2e}
\usepackage{url}
\usepackage{cite}
\usepackage{adjustbox}						
\usepackage{booktabs,array}	
\usepackage{comment}
\usepackage{cancel}
\usepackage{bm}
\usepackage{amssymb,amsthm}
\usepackage{thmtools}
\usepackage{cleveref}
\usepackage[utf8]{inputenc}
\usepackage[T1]{fontenc}
\usepackage{amsmath,amssymb,amsthm}
\usepackage{thmtools}
\declaretheoremstyle[
spaceabove=6pt, spacebelow=6pt,
headfont=\normalfont\bfseries,
notefont=\mdseries, notebraces={(}{)},
bodyfont=\normalfont,
postheadspace=0.6em,
headpunct=:
]{mystyle}
\declaretheorem[style=mystyle, name=Hypothesis, preheadhook={}]{hyp}

\makeglossaries
\newacronym{BER}{BER}{bit-error-rate}
\newacronym{AC}{AC}{alternative current}
\newacronym{AE}{AE}{autoencoder}
\newacronym{ANN}{ANN}{artificial neural network}
\newacronym{AVAR}{AVAR}{Allan variance}
\newacronym{APD}{APD}{avalanche photo detector}
\newacronym{AR}{AR}{autoregressive}
\newacronym{AWG}{AWG}{arbitrary signal generator}
\newacronym{CC}{CC}{Convolutional Code}
\newacronym{CamCom}{CamCom}{optical camera communications}
\newacronym{CAE}{CAE}{convolutional autoencoder}
\newacronym{CIR}{CIR}{channel impulse response}
\newacronym{CLT}{CLT}{Central Limit Theorem}
\newacronym{CMOS}{CMOS}{Complementary Metal Oxide Semiconductor}
\newacronym{CNN}{CNN}{Convolutional Neural Network}
\newacronym{COTS}{COTS}{commercial off-the-shelf}
\newacronym{C-V2X}{C-V2X}{cellular-vehicle-to-everything}
\newacronym{CSK}{CSK}{Color Shift Keying}
\newacronym{DAE}{DAE}{denoising auto-encoder}
\newacronym{DSRC}{DSRC}{Dedicated Short Range Communication}
\newacronym{DMT}{DMT}{discrete-multi-tone}
\newacronym{DRL}{DRL}{day time running light}
\newacronym{DPD}{DPD}{digital pre-distorter}
\newacronym{FEC}{FEC}{Forward Error Correction}
\newacronym{FLP}{FLP}{fast locking pattern}
\newacronym{FoV}{FoV}{field of view}
\newacronym{FPGA}{FPGA}{Field Programmable Gate Array}
\newacronym{fps}{fps}{frame per second}
\newacronym{IF}{IF}{intermediate frequency}
\newacronym{IM/DD}{IM/DD}{intensity modulation and direct detection}
\newacronym{ISI}{ISI}{intersymbol interference}
\newacronym{ITS}{ITS}{intelligent transportation systems}
\newacronym{KNN}{KNN}{k-nearest neighbor classifier}
\newacronym{LED}{LED}{light emitting diode}
\newacronym{LNA}{LNA}{low noise amplifier}
\newacronym{LoS}{LoS}{line-of-sight}
\newacronym{LSTM}{LSTM}{long short term memory}
\newacronym{LUT}{LUT}{look-up table}
\newacronym{LTE}{LTE}{Long Term Evolution}
\newacronym{MAC}{MAC}{medium access control}
\newacronym{ML}{ML}{machine learning}
\newacronym{MLP}{MLP}{multilayer perceptron}
\newacronym{MP}{MP}{Memory Polynomial}
\newacronym{MPC}{MPC}{multi path component}
\newacronym{MCS}{MCS}{modulation coding schemes}
\newacronym{MIMO}{MIMO}{multiple input multiple output}
\newacronym{MSE}{MSE}{mean square error}
\newacronym{NLoS}{NLoS}{non-line-of-sight}
\newacronym{NOMA}{NOMA}{non-orthogonal multiple access}
\newacronym{NN}{NNet}{neural network}
\newacronym{OCC}{OCC}{optical  camera  communications}
\newacronym{OLED}{OLED}{organic light emitting diodes}
\newacronym{OOK}{OOK}{on-off keying modulation}
\newacronym{OWC}{OWC}{Optical Wireless Communication}
\newacronym{PAM}{PAM}{pulse amplitude modulation}
\newacronym{PD}{PD}{photodetector}
\newacronym{PHR}{PHR}{physical header}
\newacronym{PHY}{PHY}{physical layer}
\newacronym{PMT}{PMT}{photomultiplier tube}
\newacronym{PPM}{PPM}{pulse position modulation}
\newacronym{PWM}{PWM}{pulse width modulation}
\newacronym{PSDU}{PSDU}{physical service data unit}
\newacronym{RBF}{RBF}{radial basis function}
\newacronym{ReLU}{ReLU}{Rectified Linear Unit}
\newacronym{RF}{RF}{radio frequency}
\newacronym{RLL}{RLL}{Run-Length Limited}
\newacronym{RMSE}{RMSE}{root mean square error}
\newacronym{RS}{RS}{Reed Solomon}
\newacronym{RSS}{RSS}{received signal strength}
\newacronym{RU}{RU}{Receiver unit}
\newacronym{SDR}{SDR}{Software-Defined Radio}
\newacronym{SHR}{SHR}{synchronization header}
\newacronym{SNR}{SNR}{signal-to-noise ratio}
\newacronym{SPAD}{SPAD}{single photon avalanche diode}
\newacronym{TDP}{TDP}{topology dependent pattern}
\newacronym{TU}{TU}{Transmitter Unit}
\newacronym{USRP}{USRP}{Universal Software Radio Peripheral}
\newacronym{V-I}{V-I}{voltage-current}
\newacronym{V2I}{V2I}{vehicle-to-infrastructure}
\newacronym{V2V}{V2V}{vehicle-to-vehicle}
\newacronym{V2X}{V2X}{vehicle-to-everything}
\newacronym{VLC}{VLC}{visible light communication}
\newacronym{V2P}{V2P}{vehicle-to-pedestrian}
\newacronym{VLP}{VLP}{visible light positioning}
\newacronym{VNA}{VNA}{vector network analyzer}
\newacronym{VPPM}{VPPM}{variable pulse position modulation}
\newacronym{DCO-OFDM}{DCO-OFDM}{direct current biased optical OFDM}
\newacronym{OFDM}{OFDM}{orthogonal frequency division multiplexing}
\newacronym{PAPR}{PAPR}{peak-to-average-power ratio}
\newacronym{PA}{PA}{power amplifier}
\newacronym{MAPE}{MAPE}{mean absolute percentage error}
\newacronym{AWGN}{AWGN}{additive white Gaussian noise}
\newacronym{FOV}{FOV}{field-of-view}
\newacronym{PSD}{PSD}{power spectral density}
\newacronym{IFFT}{IFFT}{inverse fast Fourier transform}
\newacronym{FFT}{FFT}{fast Fourier transform}
\newacronym{VVLC}{V-VLC}{vehicular visible light communications}

\def\BibTeX{{\rm B\kern-.05em{\sc i\kern-.025em b}\kern-.08em
    T\kern-.1667em\lower.7ex\hbox{E}\kern-.125emX}}

\begin{document}

\title{Vehicular Visible Light Communications \\ Noise Analysis and  Autoencoder Based Denoising}

\author{\IEEEauthorblockN{Bugra Turan$^1$, O. Nuri Koc$^2$, Emrah Kar$^2$ and Sinem Coleri$^1$}\\
\IEEEauthorblockA{$^1$Department of Electrical and Electronics Engineering, Koc University, Sariyer, Istanbul, Turkey, 34450\\
E-mail: bturan14@ku.edu.tr, scoleri@ku.edu.tr \\
$^2$Koc University Ford Otosan Automotive Technologies Laboratory, Sariyer, Istanbul, 34450, Turkey.\\
E-mail: okoc@ku.edu.tr, ekar@ku.edu.tr}}


\maketitle

\begin{abstract}

\Ac{VVLC} is a promising \ac{ITS} technology for \ac{V2V} and \ac{V2I} communications with the utilization of \glspl{LED}. The main degrading factor for the performance of \ac{VVLC} systems is noise. Unlike traditional \ac{RF} based systems, \ac{VVLC} systems include many noise sources: solar radiation, background lighting from vehicle, street, parking garage and tunnel lights. Traditional \ac{VVLC} system noise modeling is based on the additive white Gaussian noise assumption in the form of shot and thermal noise. In this paper, to investigate both time correlated and white noise components of the \ac{VVLC} channel, we propose a noise analysis based on \ac{AVAR}, which provides a time-series analysis method to identify noise from the data. We also propose a generalized Wiener process based \ac{VVLC} channel noise synthesis methodology to generate different noise components. We further propose \ac{CAE} based denoising scheme to reduce \ac{VVLC} signal noise, which achieves reconstruction \ac{RMSE} of 0.0442 and 0.0474 for indoor and outdoor channels, respectively.   


\end{abstract}




\IEEEpeerreviewmaketitle

\section{Introduction}

Vehicular communications, enabling various form of connectivity between road users such as \ac{V2V}, \ac{V2I} and \ac{V2P} communications, aim to increase road safety and efficiency through the exchange of traffic, road and vehicle information. Currently, \ac{RF} based \ac{C-V2X} \cite{molina2017lte} and \ac{DSRC} \cite{arena2020review} technologies are standardized to support \ac{V2X} applications. However, omnipresence of \glspl{LED} in the vehicle lighting paves the way for \ac{VVLC} as a complementary technology to the existing schemes. Since \ac{VVLC} provides \ac{RF} interference, jam and spoof free directional optical communications, it is mainly foreseen to be a strong candidate to enable secure vehicular communications.

\ac{VVLC} signal quality substantially degrades with the noise due to ambient lighting sourced from solar radiation, traffic lights, road lighting, and vehicle lights in the \ac{VVLC} channel. Thus, accurate noise characterization and artificial noise synthesis plays an important role in the evaluation of system performance to address different background lighting conditions. Moreover, reducing \ac{VVLC} signal noise increases \ac{SNR}, increasing symbol detection probability. 

Noise for the \ac{VLC} schemes is generally assumed to composed of shot noise and thermal noise in the form of white Gaussian noise \cite{ghassemlooy2019optical}. Shot noise arises due to photons incident on the \ac{PD} active area and increases with the square root of the incident optical power. Therefore, background lights, together with the modulated \ac{VVLC} transmitter are the main sources of shot noise. On the other hand, thermal noise is mainly generated by the receiver circuits, due to irregular motion of electrons in conducting materials \cite{komine2004fundamental}. Thermal noise increases with the operating temperature of receiver electronics and decreases with the increasing resistance, which is independent from the incident optical power. In the literature, \ac{VLC} noise is characterized in \cite{hua2018noise} for indoor environments, where the noise is demonstrated to exhibit time correlated characteristics due to \ac{LED} lighting. Moreover, noise for outdoor \ac{VVLC} is investigated in \cite{islim2017investigation, farahneh2018performance}, where \cite{islim2017investigation} presents the \ac{VLC} system \ac{BER} degradation due to solar irradiance induced noise and \cite{farahneh2018performance} provides \ac{VVLC} \ac{BER} analysis with the consideration of thermal, shot, dark and \ac{ISI} noise. However, none of the studies to date has empirically characterized the \ac{VVLC} noise in terms of different noise components for various ambient light sources. Moreover, \ac{VVLC} signal denoising has not been investigated to date. 

In this paper, we investigate the time correlation properties of \ac{VVLC} noise due to background lighting and identify noise components by using \ac{AVAR}. We further propose a generalized Wiener process to synthesize \ac{VVLC} channel noise for simulation purposes. Moreover, we present a denoising scheme to reduce \ac{VVLC} signal noise. The main novel contributions of this paper are threefold. 
\begin{itemize}
    \item We analyze the time correlation of noise together with \ac{LED} transmitter-optical receiver distance dependence in the \ac{VVLC} channel based on the real world data collected for solar radiation and \ac{LED} ambient light disturbed channels, for the first time in the literature.  
    \item We identify the \ac{VVLC} channel noise types by using \ac{AVAR} and artificially generate the time-correlated artificial light induced noise with a Wiener filter based recursive algorithm.
    \item We propose a \ac{CAE} based \ac{VVLC} signal denosing scheme to remove the adversarial effects of ambient light induced noise, for the first time in the literature. 
\end{itemize}

The rest of the paper is organized as follows. Section~\ref{sec:method} provides the methodologies used to identify noise components, synthesize noise and denoising. Section~\ref{sec:meas} details the experimental setup which is used to capture noise and \ac{VVLC} signals for different ambient lighting conditions. In Section~\ref{sec:results}, we provide the results for noise characterization, noise synthesis and the proposed denoising scheme performance. Section~\ref{sec:conc} concludes the paper. 


\noindent

\section{Methodology} \label{sec:method}

\subsection{Noise Analysis}\label{sec:noiseanalysis}

\ac{VVLC} receivers generate electrical signals from the incident photons on \ac{PD}. The detection of photons from the \ac{VVLC} transmitter and background lighting is considered to be the primary noise source for \ac{VVLC}, known as photon shot noise. Moreover, the variation in the amount of \ac{PD} dark current due to the thermal excitation of electrons within the silicon chip in the absence of incidence photons constitutes the dark shot noise. On the other hand, the optical-to-electrical converted signals are amplified at the receiver by trans-impedance circuits, which generate thermal noise by the thermal fluctuation of electrons in the load resistance of the circuits. The shot noise and thermal noise are generally assumed to follow Gaussian distribution, which yields limited accuracy for noise modeling, since various forms of ambient lights can lead different noise distributions. 

Considering a \ac{VVLC} transceiver with a \ac{LED} transmitter and optical receiver, the \ac{VVLC} received signal is described as 

\begin{equation}
y(t) = R  x(t)   \otimes {h}(t) + v_n(t), \label{tr_model}
\end{equation}

\noindent where x(t) is the transmitted \ac{VVLC} signal, $R$ is the responsivity of \ac{PD} (A/W), $h(t)$ is the optical channel response, and $v_n(t)$ represents the channel noise. 

The $v_n(t)$ can be in the form of white noise, where the values at any point are independent and uncorrelated from the other values, or colored noise, where the values are correlated to the other points. The noise is characterized by its power spectrum, which decays as $1/f^{\alpha}$, where $\alpha \geq 0$ describes the type of noise (i.e., $\alpha=0$ white noise, $\alpha=1$ flicker noise, $\alpha=2$ random walk noise) and $f$ represents the frequency \cite{kasdin1995discrete}. In the context of \ac{VLC}, white noise, flicker noise and random walk noise can be regarded as dominant noise sources due to receiver electronics, DC offset and \ac{LED} modulation, respectively \cite{hua2018noise}.

When no background lighting is present, $v_n(t)$ can be modeled as a signal independent, zero mean Gaussian random variable with standard deviation of $\sigma$ and \ac{PSD} given by, $    S_{w}(f)=\frac{N_{0}}{2}$, where $N_{0}=2\sigma^2$. However, considering background lighting from other light sources such as street lights, and other vehicle lights, noise can exhibit time correlated behavior at different correlation times \cite{hua2018noise}. Therefore, time correlation of \ac{VLC} noise for different channels is investigated by using Ljung Box test to identify the noise time correlation characteristics, where the identified noise components are further quantified with the noise coefficients by using \ac{AVAR}.


\paragraph{Ljung Box Q-test}  
The Ljung-Box method tests the auto-correlation structure of the data samples for the null hypothesis of independent distribution of the data against the hypothesis of the correlated data by using test statistic $Q_m$ \cite{ljung1978measure}. 
Let the channel noise process with $N$ samples be defined as $X_{\sigma}(j), j=1,...,N$, then, the sample auto-correlation, $\widehat{p}(k)$, at lag $k$, is given by
\begin{equation}
 \widehat{p}(k)=\frac{\sum_{j=k+1}^{N}X_\sigma (j)X_\sigma(j-k)}{\sum_{j=1}^{N}X_\sigma^2(j) }\qquad k\in\mathbb{Z}
\end{equation}


\noindent The  null and alternative hypotheses are 

 \setcounter{hyp}{-1}
 \begin{hyp}[White noise] \label{hyp:a} $\widehat{p}(k)=0 \qquad \forall k\neq0$ \end{hyp}
 \begin{hyp} [Colored noise] \label{hyp:b} $\widehat{p}(k)\neq0 \quad \text{for some} \hspace{0.5em} k\neq0. $ \end{hyp}


Considering the null hypothesis, the Ljung-Box test statistic, $Q_m$, is calculated as

\begin{equation}
 Q_m=N(N+2)\sum_{j=1}^{m}\frac{\widehat{p}^2(k)}{N-k}
\end{equation} 

\noindent where $m$ is  the  number  of  lags. 
Considering the significance level $\alpha$, the threshold above which the the null hypothesis is rejected becomes the $\alpha$-quantile as $Q_{m} > \chi_{1-\alpha,m}^2$, where $\chi_{1-\alpha,m}^2$ is the value of the chi-square distribution corresponding to the significance level $\alpha$ and $m$ degrees of freedom. 

To further analyze noise characteristics by the identification of noise coefficients for white, flicker and random walk noise, \ac{AVAR} method is used. 

\paragraph{Allan Variance} \ac{AVAR} provides a time-series analysis method to identify noise from the data based on the derivation of the evolution of root mean squares random drift error with time. To obtain the Allan variance for a discrete random variable $X_k$ with $k=1,2,...,N$ samples, first, clusters from $n$ samples are formed, where $n<N/2$, and the total number of clusters is the integer of $\frac{N}{n}$. Then, the mean value of two adjacent clusters is calculated as

\begin{equation}
    \bar{X}_k(n)=\frac{1}{n}\sum_{i=k}^{k+n-1}X_i
\end{equation}
\begin{equation}
    \bar{X}_{k+n}(n)=\frac{1}{n}\sum_{i=k+n}^{k+2n-1}X_i
\end{equation}



The \ac{AVAR} of cluster with length $\tau$ can be described as

\begin{equation}
\label{eq:AVAR}
    \sigma ^2(\tau )=\frac{1}{2(N-2n+1)}\sum_{k=1}^{N-2n+1}\left (\bar{X}_{k+n}(n)-\bar{X}_{k}(n)\right )^2.
\end{equation}

\noindent where the log-log plot of \ac{AVAR} calculated by (\ref{eq:AVAR}) with cluster length allows to identify different noise types by the slope of the curves, where the curve slope of the Allan deviation (i.e., square root of \ac{AVAR}), $\sigma(\tau)$ is $-1/2$, $0$ and $1/2$ for white, flicker and random walk noise, respectively \cite{ieee2006ieee}. 

Considering two-sided \ac{PSD}, $S_{x}(f)$, of the continuous random process, the relationship between the \ac{AVAR} and the  $S_{x}(f)$ is given by \cite{hua2018noise} 

\begin{equation}
    \sigma ^2(T)= 4\int_{0}^{\infty}S_{\rm X}(f)\frac{sin^4(\pi fT)}{(\pi fT)^2}d_{\rm f}
 \end{equation}

\noindent where substitution of the considered noise \ac{PSD} into the equation and performing integration yields noise coefficients as defined in the IEEE standard \cite{ieee2006ieee}.

Time-correlated random walk noise is observed for \ac{VLC} channel due to Brownian motion in the receiver circuit \cite{hua2018noise}, which is given by $\sigma^2(\tau)=\frac{K^{2}\tau}{3}$, where $K$ is the random walk coefficient, calculated by $K=\sigma(3)$. Flicker noise, observed in almost all electronic devices including \ac{VLC} receivers due to the flow of direct current is given by, $\sigma^2(\tau)=\frac{2B^{2} ln2}{\pi}$, where $B$ is the flicker noise coefficient, represented by $B=\sigma (\sqrt3)$. Therefore, the \ac{VVLC} system channel noise is identified by the extraction of flicker noise ($B$), white noise ($N$) and random-walk ($K$) coefficients evaluating the log-log plots of  $\sigma(\tau)$, and $\tau$ at the respective values (i.e., white noise =$\sigma$(1), flicker = $\sigma$(3), random walk = $\sigma$($\sqrt{3}$).  

\subsection{Noise Synthesis}

Noise synthesis is useful to simulate practical \ac{VVLC} noise behaviour for different background lighting conditions, where white noise and correlated perturbations in the form of flicker noise and random walk can be obtained by using \ac{AVAR}. Then, a recursive Wiener filter based method is employed to generate artificial \ac{VVLC} noise yielding the desired noise coefficients. The colored noise is defined as the convolution of white noise ($w \sim \mathcal{N}(\mu,\,\sigma^{2})$) with an impulse response function, where a white noise process is filtered by using a generalized Wiener filter for $\alpha \in (0,2)$ as

\begin{equation}
H^{\alpha}(z)=\frac{1}{(1-z^{-1})^{\alpha/2}}
\end{equation}

Representing $H^{\alpha}(z)$ as the power series, $H^{\alpha}(z)=\sum_{j=0}^{\infty} H^{\alpha}_{j} z^{-j} $, the weights are recursively calculated by \cite{kasdin1995discrete}

\begin{equation}
\begin{gathered}
    H_{0}^{\alpha}=1\\
    H_{j}^{\alpha}=\left (  \frac{\alpha}{2}+j-1\right )\frac{H_{j-1}^{\alpha}}{j}
    \end{gathered}
\end{equation}
Then, taking the inverse Z transform of $H^{\alpha}(z)$ yielding $h_{i}^{\alpha}$ and substituting into the discrete convolution, the colored noise vector of length $M$ is obtained by 

\begin{equation}
    \eta^{(\alpha,M)}=\sum_{j=0}^{i} H^{\alpha}_{j} w_{i-j} \quad \textrm{for} \quad i=0,..,M-1
    \end{equation}

\noindent where $w_{j}$ is an independent and identically distributed white noise sequence. The model enables white, flicker and random walk noise samples generation with the selection of $\sigma^{2}$ and $\alpha$.

\subsection{Denoising}

In practical \ac{VVLC} scenarios, noise contaminated \ac{VVLC} symbols lead to lower \ac{SNR} and higher \ac{BER} due to erroneous detection of the symbols. Moreover, exact characterization, identification and subtraction of the noise requires a comprehensive analysis, since noise sources vary for mobile scenarios. Thus, we propose a \ac{CAE} based denoising scheme, which learns the noise characteristics from the training data to reconstruct the pulse based \ac{VVLC} symbols from their noisy counterparts.

The \ac{CAE} combines \glspl{CNN} \cite{goodfellow2016deep} with \ac{AE} \cite{vincent2010stacked}, where \ac{CNN} algorithm takes the input data and assigns importance to the various aspects of the data in form of weights and biases enabling differentiation of the features. For \ac{CNN} each layer is connected to another with a limited number of connections, and the same pattern is used for all subsequent layer connections yielding special down-sampling (pooling) layers. Thus, the search space for the number of parameters is reduced. \ac{CNN} consists of convolutional layers, which include a set of filters to extract feature maps describing the input data; max-pooling layers to obtain translation invariance and dimension reduction with down-sampling; and a fully connected layer, which provides regression or classification. On the other hand, \ac{AE} is a multilayered \ac{ANN}, which aims to replicate the input at the output and learns by supervision. \ac{AE} consists of input, encoding and decoding layers. At the encoder, noise corrupted input data $\hat x$ is mapped to a hidden representation $z$ by using a non-linear transformation given by $z=f(Wx+b)$, whereas the hidden representation is mapped back to reconstructed data $\hat x$ at the decoder through another non-linear transformation as $\hat x= g(\hat W z+ \hat b)$, where $f$ and $g$ are non-linear activation functions such as \ac{ReLU}, $W$ and $b$ represent the weight and bias matrix of the encoder, $\hat W$ and $\hat b$ represent the weight and bias matrix of the decoder, respectively. $W$, $b$, $\hat W$ and $\hat b$ are minimized with the loss function.  




\section{Measurement Setup} \label{sec:meas}
To analyze the \ac{VVLC} channel noise time dependence characteristics due to background lighting sources, we test three different data-sets where the channel is disturbed by indoor parking garage \ac{LED} lighting, outdoor solar radiation, and vehicle \ac{LED} lights. At the receiver side, a \ac{PD} (Thorlabs-PDA100A) with a plano convex lens (Thorlabs-LA1050-A) converts the ambient light optical signals into electrical signals. Then, a high speed digital storage oscilloscope (Rohde$\&$Schwarz RTA-4004) is used to digitize and store the measurement samples. The background lighting of the indoor parking garage is composed of 8 \glspl{LED} driven by \ac{AC}, whereas vehicle (Fiat Tipo 2018) \ac{LED} \ac{DRL} light is driven by a \ac{PWM} circuit to enable dimming control while ensuring power efficiency. On the other hand, solar radiation is considered as the main noise source for outdoor background light during day conditions, where noise due to other artificial light sources (i.e. vehicle \ac{DRL} \ac{LED} lights) is not considered. To analyze \ac{VVLC} noise, N=1000000 measurement samples are recorded with 625 MSPS sample rate. 

To train and test the \ac{CAE} denoising scheme, \ac{OOK} based \ac{VVLC} signals with 1 ms ON and 4 ms OFF period are recorded at 10 m distance from a single vehicle \ac{DRL} \ac{LED} light. For indoor parking garage scenario, 8 \glspl{LED} are turned on, whereas for outdoor scenarios, bright sun and shadow conditions without any artificial noise are considered to address different noise levels. For each record, N=4000000 samples with 1.67 MSPS sample rate, corresponding to 2.39 $s$ time length, are recorded. Considering ON and OFF periods of the \ac{VVLC} sequences, each record consists of approximately 500 pulses. The noise free pulses are recorded in an indoor parking garage with complete darkness. For each noise condition, 15 measurement runs are executed to form the training and test data sets of 5000 pulse samples per scenario.

\section{Results and Analytics}\label{sec:results}

\subsection{Noise Characterization}

\begin{figure}
\centering
\includegraphics[clip,trim=0 0 0 20,width=1\linewidth]{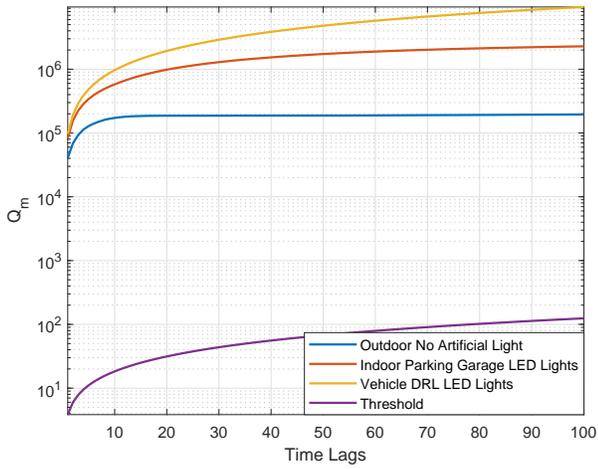}
\caption{Ljung-Box Q-test for 100 time lags}
\label{Fig:ljung}
\end{figure}

To investigate the time dependence of the \ac{VVLC} channel noise samples, Ljung-Box Q-test is performed on measurements samples from three different noise disturbed \ac{VVLC} channels. 

Fig.~\ref{Fig:ljung} shows the Ljung-Box Q statistics values for $m = 1, 2, . . . , 100$ with the threshold statistics to reject the null hypothesis at $\alpha$ = 0.05 significance level. The obtained statistic values are far above the threshold, yielding lower $p$-values below $\alpha$ which indicates that the null hypothesis can be rejected. Therefore, the test results indicate that the \ac{VVLC} channel noise exhibits a time dependent statistical behaviour. Moreover, vehicle \ac{LED} disturbed channel indicates the highest time correlation, whereas the outdoor \ac{VVLC} channel under solar radiation has the lowest time correlation. To further identify the noise characteristics of the measurement samples with the noise coefficients, \ac{AVAR} method is employed. 


\begin{table}[]
\caption{Scenario Based  Noise Coefficients}
\begin{adjustbox}{width=1\linewidth}
\begin{tabular}{|l||c|c|c|}
\hline
\multirow{2}{*}{\textbf{Channel}}                               & \multicolumn{3}{c|}{\textbf{Noise Coefficient}}          \\ \cline{2-4} 
                                                               & Random Walk (K) & Flicker Noise (B) & White Noise (N) \\ \hline
\begin{tabular}[c]{@{}l@{}}Indoor Garage\\ 8 LEDs\end{tabular} & 0.0117          & $7.49\times 10^{-5} $           & $ 8.78\times 10^{-8} $         \\ \hline
\begin{tabular}[c]{@{}l@{}}Indoor Garage\\ 1 LED\end{tabular}  & 0.0022          & $4.09\times 10^{-5} $           & $1.10\times 10^{-8}   $       \\ \hline
Outdoor Solar Radiation                                         & N/A             & $1.27\times 10^{-5} $           & $1.99\times 10^{-8}   $       \\ \hline
Vehicle DRL LED                                               & 0.1129          & 0.0167            & $4.08\times 10^{-8}   $       \\ \hline
\end{tabular}
\end{adjustbox}
\label{tab:coeffs}
\end{table}

\begin{figure} 
    \centering
  \subfloat[\label{Fig:allan}]{%
       \includegraphics[clip,trim=0 0 0 20,width=0.5\linewidth]{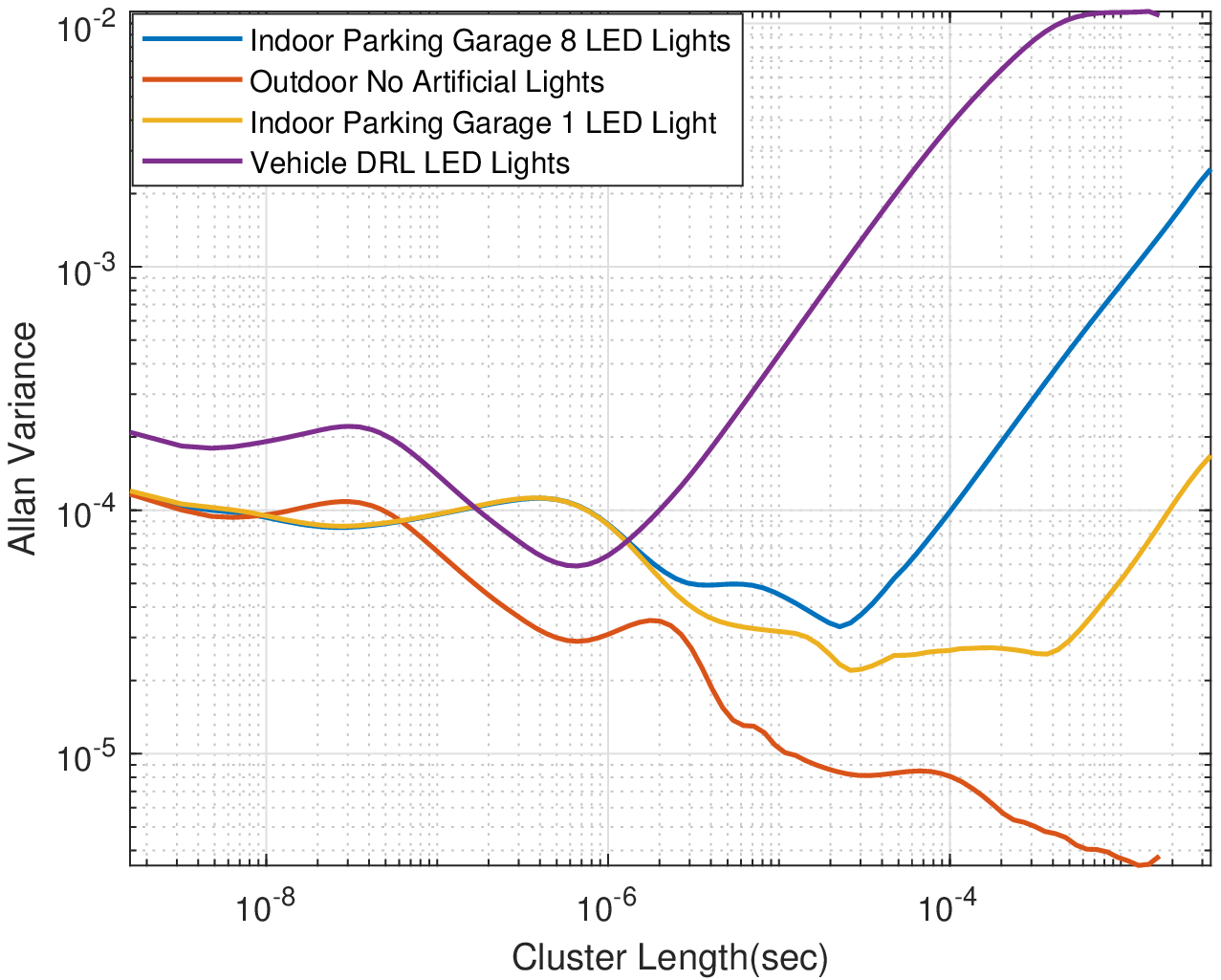}}
    \hfill
  \subfloat[\label{Fig:allandistance}]{%
        \includegraphics[clip,trim=0 0 0 20,width=0.5\linewidth]{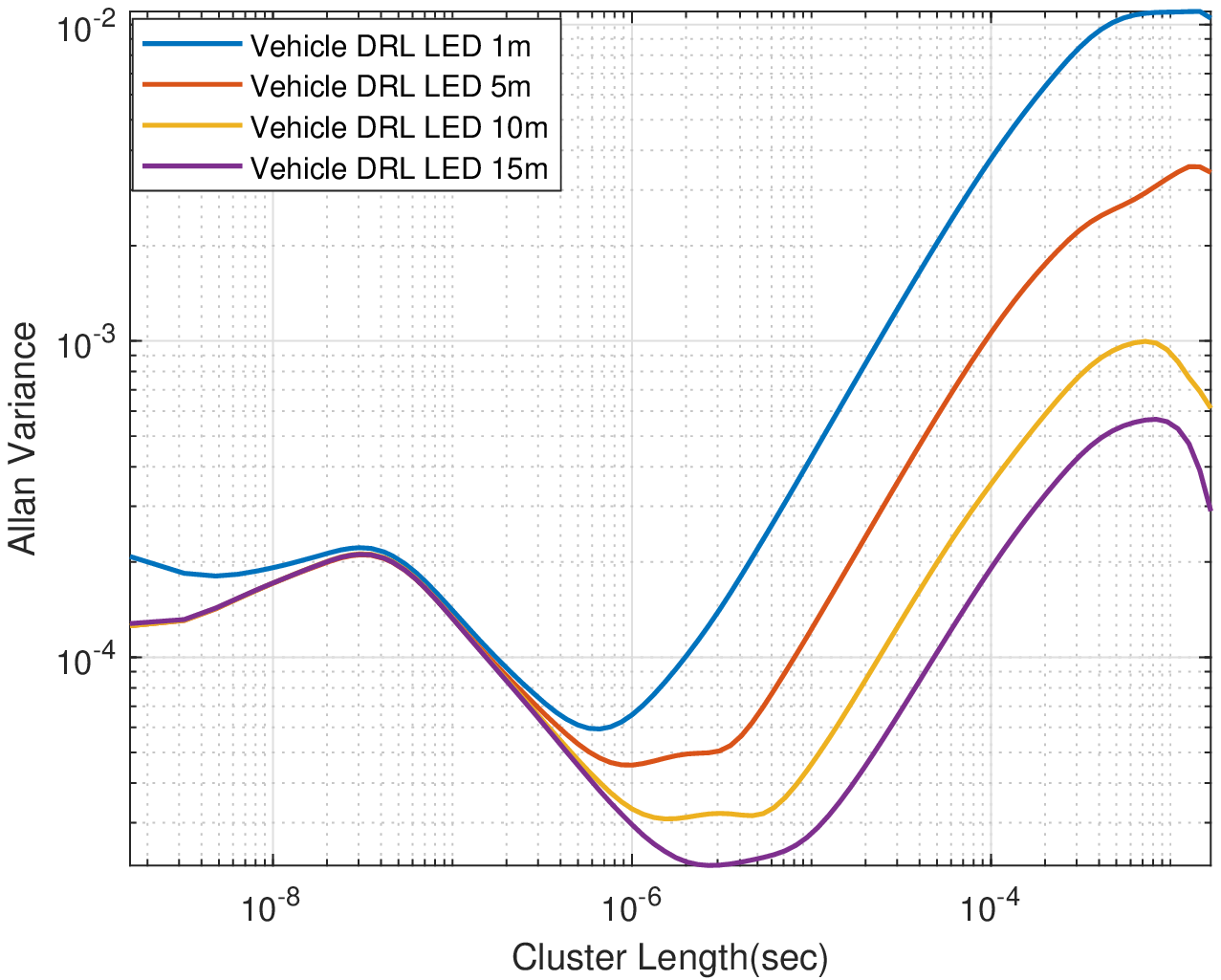}}
       \\
  \caption{(a) Allan variance analysis for different \ac{VVLC} channels and (b) transmitter-receiver distances}
  \label{fig:allananalysis}
\end{figure}

Fig.~\ref{fig:allananalysis}~(a) demonstrates the \ac{AVAR} for the scenarios under consideration. The curves for outdoor and ambient light free dark indoor scenarios show flicker noise and white noise characteristics, whereas illuminated indoor parking garage and vehicle \ac{LED} light disturbed \ac{VVLC} channels exhibit flicker noise and white noise along with the random walk noise. The noise coefficients for the considered scenarios are listed in Table~\ref{tab:coeffs}. The random walk coefficient is the highest for the \ac{PWM} modulated vehicle \ac{LED} disturbed channel among all scenarios whereas increased number of \glspl{LED} is shown to increase the random walk for indoor scenarios. Moreover, the white noise coefficient for indoor \ac{LED} light disturbed channel is slightly higher when compared to non-illuminated indoor channel, where \ac{AC} power driven \glspl{LED} are observed to increase the white noise at the receiver. On the other hand, flicker noise increases with the DC offset of the illumination, where the vehicle \ac{LED} directed towards the receiver generates the highest DC offset together with the strongest flicker noise. 


\begin{table}[]
\centering
\caption{Distance Dependent Noise Coefficients for Vehicle LED}
\begin{adjustbox}{width=1\linewidth}
\begin{tabular}{|l||c|c|c|}
\hline
\multicolumn{1}{|c||}{\multirow{2}{*}{\textbf{Scenario}}} & \multicolumn{3}{c|}{\textbf{Noise Coefficient}}           \\ \cline{2-4} 
\multicolumn{1}{|c||}{}                                   & Random Walk (K) & Flicker Noise (B) & White Noise (N) \\ \hline
Vehicle LED DRL 5 m                                      & 0.2195          & 0.0053            & $4.41 \times 10^{-8}$        \\ \hline
Vehicle LED DRL 15 m                                     & 0.0152          & $3.30\times 10^{-5}$            & $2.94\times 10^{-8}  $        \\ \hline
\end{tabular}
\end{adjustbox}
\label{tab:distancecoeffs}
\end{table}

Fig.~\ref{fig:allananalysis}~(b) depicts the distance dependent noise characteristics of the vehicle \ac{LED} disturbed \ac{VVLC} channel and Table~\ref{tab:distancecoeffs} shows the noise coefficients at varying distances. Both the white noise and random walk coefficients decrease with the increasing distance. However, random walk noise decreases substantially when compared to white noise. Thus, the fundamental noise due to \ac{PWM} modulated vehicle \ac{LED} lights is random walk, where the main source of the white noise can be regarded as thermal noise, independent of incident light.    

\begin{figure*} 
    \centering
  \subfloat[\label{synthWN}]{%
       \includegraphics[clip,trim=10 0 30 20,width=0.33\linewidth]{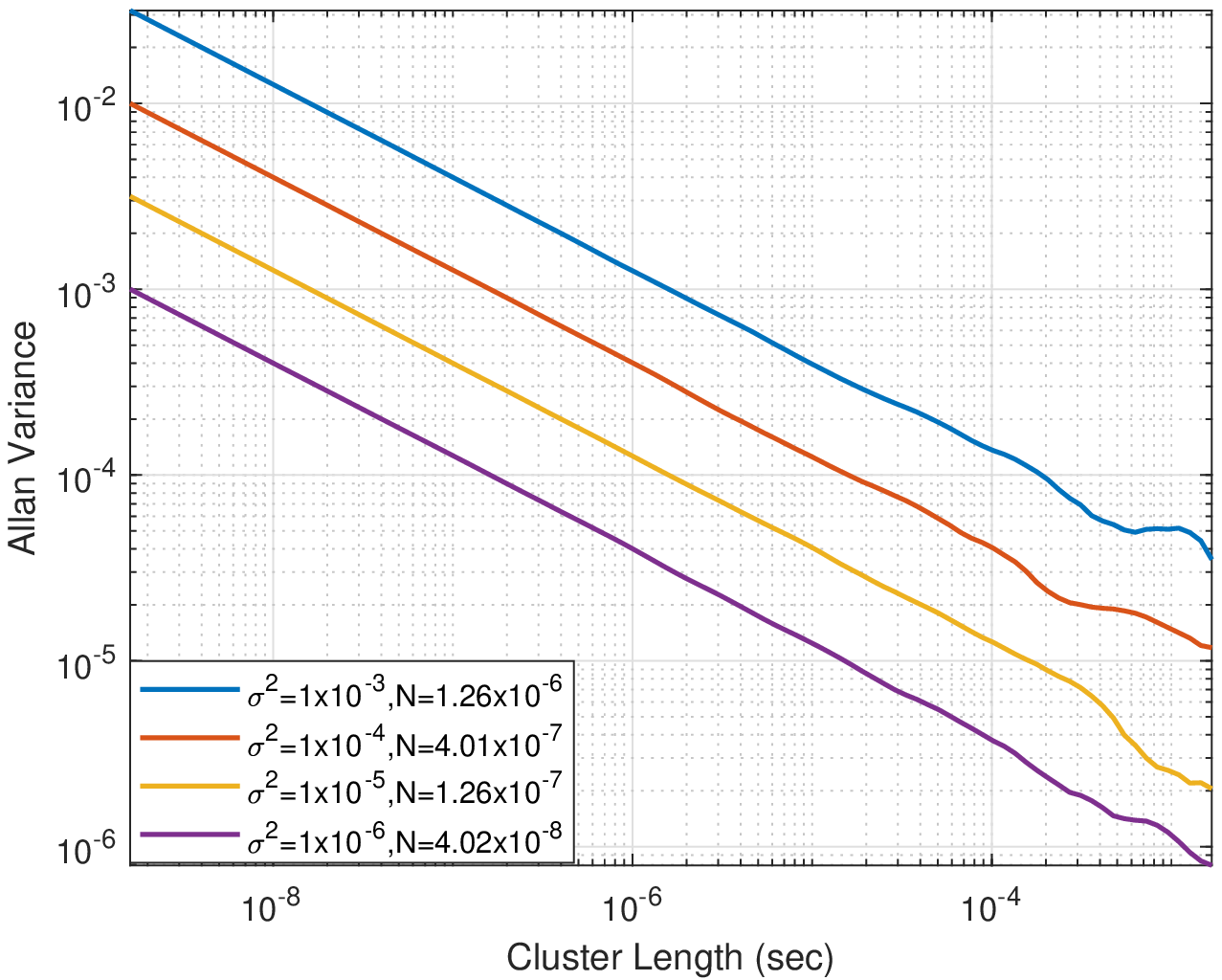}}
    \hfill
  \subfloat[\label{synthB}]{%
        \includegraphics[clip,trim=10 0 30 20,width=0.33\linewidth]{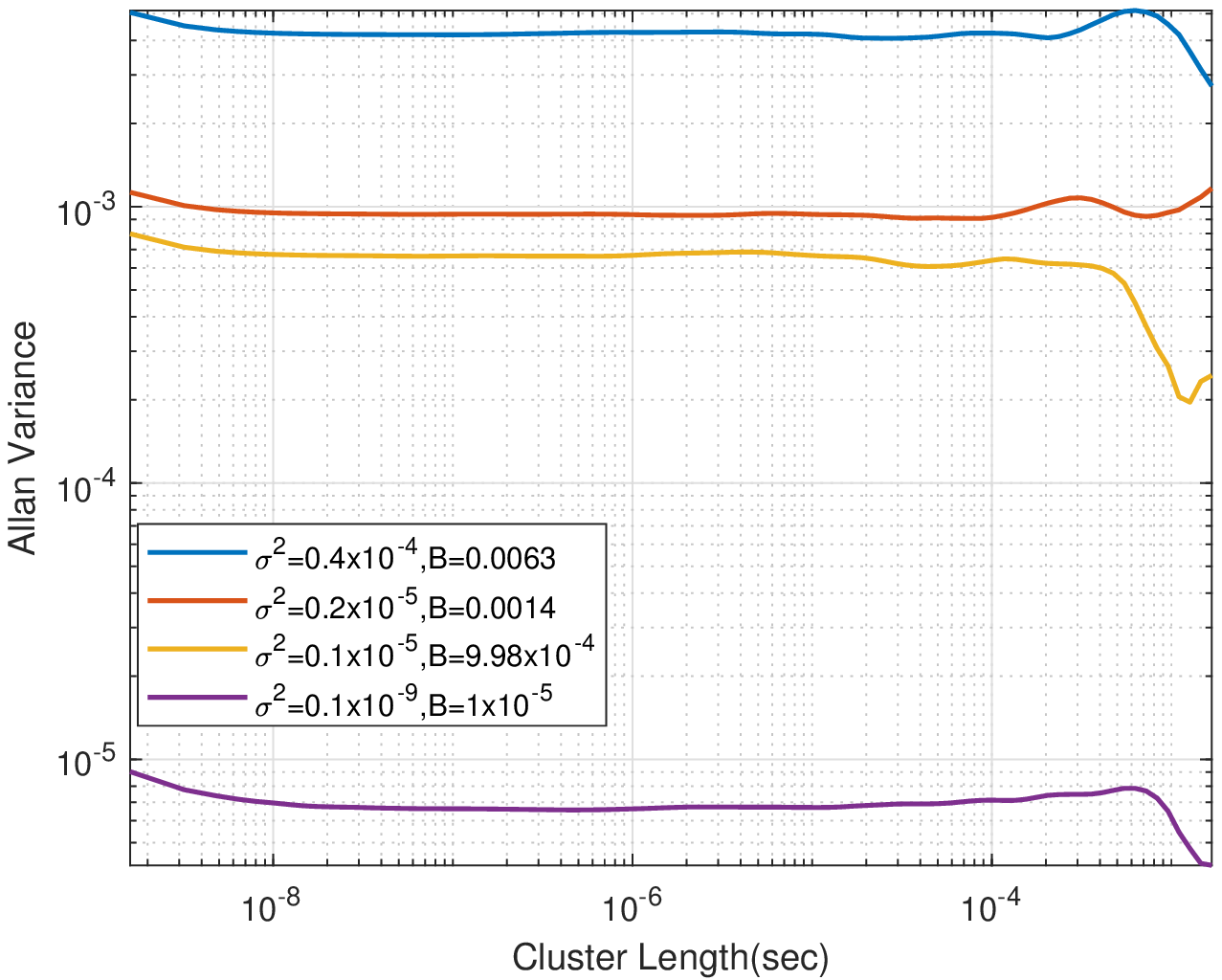}}
         \hfill
  \subfloat[\label{synthRV}]{%
        \includegraphics[clip,trim=10 0 30 20,width=0.33\linewidth]{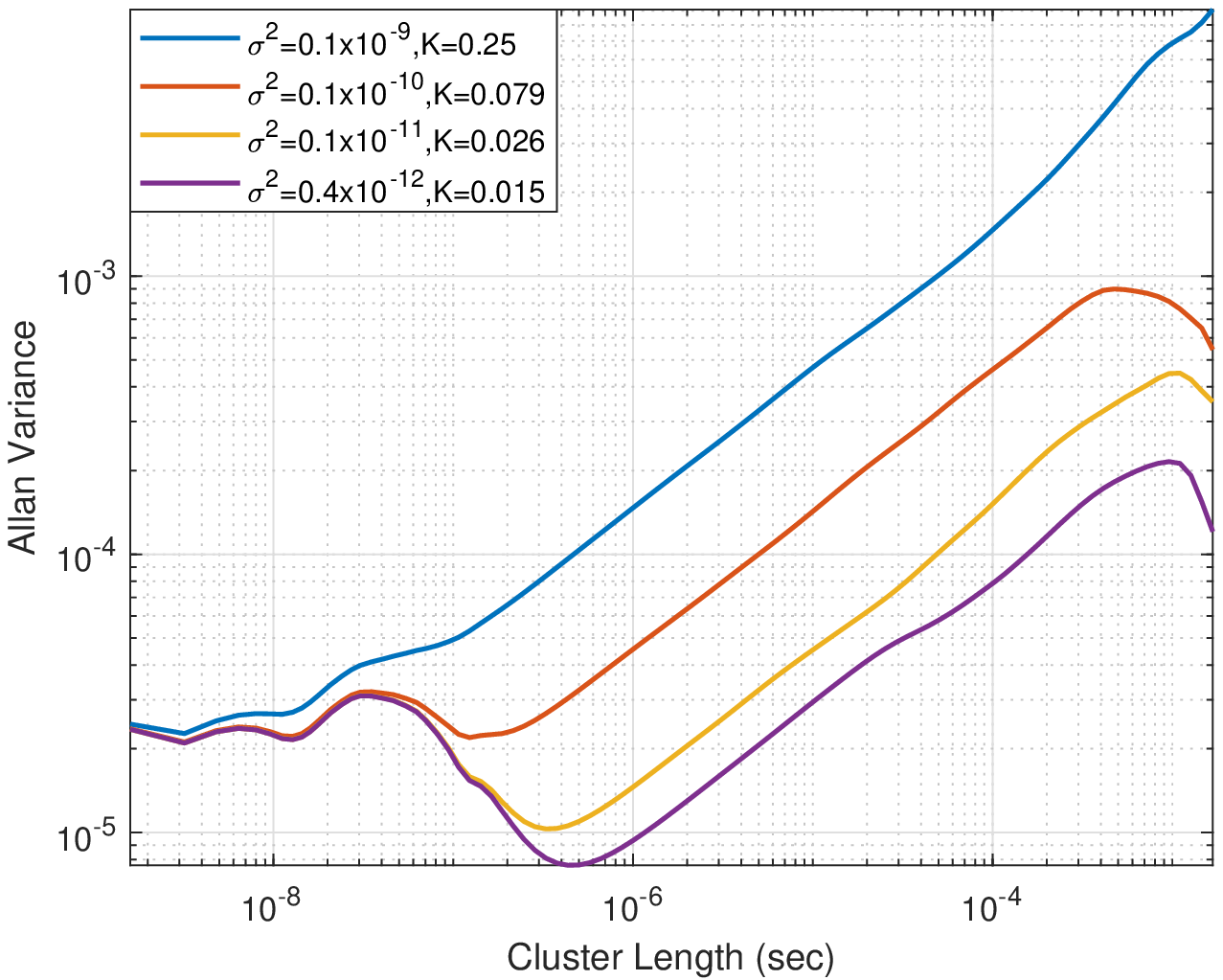}}
    \\
  \caption{(a) Synthesized white noise (b) flicker noise and (c) random walk noise with different $\sigma^{2}$ values}
  \label{fig:synth}
\end{figure*}

\subsection{Noise Synthesis}

The noise analysis results demonstrate the existence of colored noise along with the white noise for \ac{VVLC} channels. While the traditional white noise assumption can provide a basis for system simulations, incorporating colored noise components increases the evaluation accuracy, specifically for time varying \ac{VVLC} channels. Therefore, we are interested in simulating different noise components with varying noise coefficients. The proposed generalized Wiener process based method generates various noise samples with the selection of $\alpha$ and $\sigma^{2}$ parameters, to define the noise type and strength, respectively.  
 
Fig.~\ref{fig:synth} shows the \ac{AVAR} for synthesized white ($\alpha=0$), flicker ($\alpha=1$) and random walk ($\alpha=2$) noise with varying $\sigma^{2}$ values. Fig.~\ref{fig:synth}~(a) depicts \ac{AVAR} for the white noise disturbed channel, Fig.~\ref{fig:synth}~(b) denotes the dominant flicker noise, whereas Fig.~\ref{fig:synth}~(c) presents the combination of white, flicker and random walk noise with the strong random walk noise component. Since white noise is demonstrated to decrease with the decreasing illumination, lower values of $\sigma^2$ can be used to simulate low illumination scenarios yielding smaller $N$ coefficients. On the other hand, stronger flicker noise due to DC offset can be simulated with increasing $\sigma^2$ values. Moreover, distance dependent vehicle \ac{LED} disturbed channel noise can be simulated with the algorithm as depicted in Fig.~\ref{fig:synth}(c) where the decreasing $K$ with increasing distance can be simulated by the decreasing values of $\sigma^2$. Therefore, the recursive noise synthesis algorithm can be used to generate artificial \ac{VVLC} noise datasets.

\subsection{Denoising}

The proposed \ac{CAE} denoiser is trained with the database of noisy and clean signals. The database consists of 5.000 \ac{VVLC} \ac{OOK} samples for indoor, outdoor bright sun and shadow \ac{VVLC} channels together with the clean signals. Three different databases, mixed data (MD), outdoor data (OD), and indoor data (ID), are formed to train and evaluate the performance of the proposed scheme. The MD data-set consists of indoor \ac{LED} light contaminated and outdoor solar radiation disturbed samples, OD data-set includes pure outdoor solar radiation disturbed bright sun and shadow scenario samples, and ID dat-set includes pure \ac{LED} light contaminated samples. The \ac{SNR} levels of the noisy signals vary due to DC offset. Therefore, all signal amplitudes are normalized between 0 and 1. The training database is divided into two subsets for training and test with 70$\%$ and 30$\%$ ratios, respectively.

The denoising scheme is implemented in Python using $Keras$ library \cite{chollet2015keras} with tensorflow \cite{abadi2016tensorflow} platform. The model is trained by using Adam optimizer \cite{kingma2014adam} with binary-cross-entropy loss function \cite{shore1980axiomatic} for 100 epochs using batch size 50 on an embedded computer, NVIDIA Jetson TX2.

In the proposed \ac{VVLC} \ac{CAE} denoising scheme, the input noisy \ac{VVLC} signals are encoded into low dimensional features at the encoder with two convolutional layers, whereas the decoder reconstructs the output from the low dimensional features with two transposed convolutional layers. In the encoder, the input signals with $2317 \times 1$ size are taken as input, then a convolutional process with 128 filters of size $3\times1$ and stride of 2 including maximum norm regularization is applied on the first layer. The next convolutional layer is made up of 32 filters with size $3\times1$ and a stride of 2. The decoder part is the symmetric inverse of the encoder, whereas at the output layer, a deconvolutional layer with one filter employs padding to yield the same size with the input data.

The \ac{RMSE} metric is used to evaluate the variance between the model output of denoised signals and the clean signals. The smaller \ac{RMSE} value indicates the better denoising performance, and \ac{RMSE} can be calculated by $ \textrm{RMSE}=\sqrt {\frac{1}{N} \times \sum_{n=1}^{N} (\bm{x_{i}}- \bm{\hat x_{i}})^2 }$, where $N$ is the number of samples, $\bm{x_{i}}$ is the clean signal vector and $\bm{\hat x_{i}}$ denotes the denoised signal vector. 




\begin{figure} 
    \centering
  \subfloat[\label{denoiseBARS}]{%
       \includegraphics[clip,trim=0 0 0 20,width=0.5\linewidth]{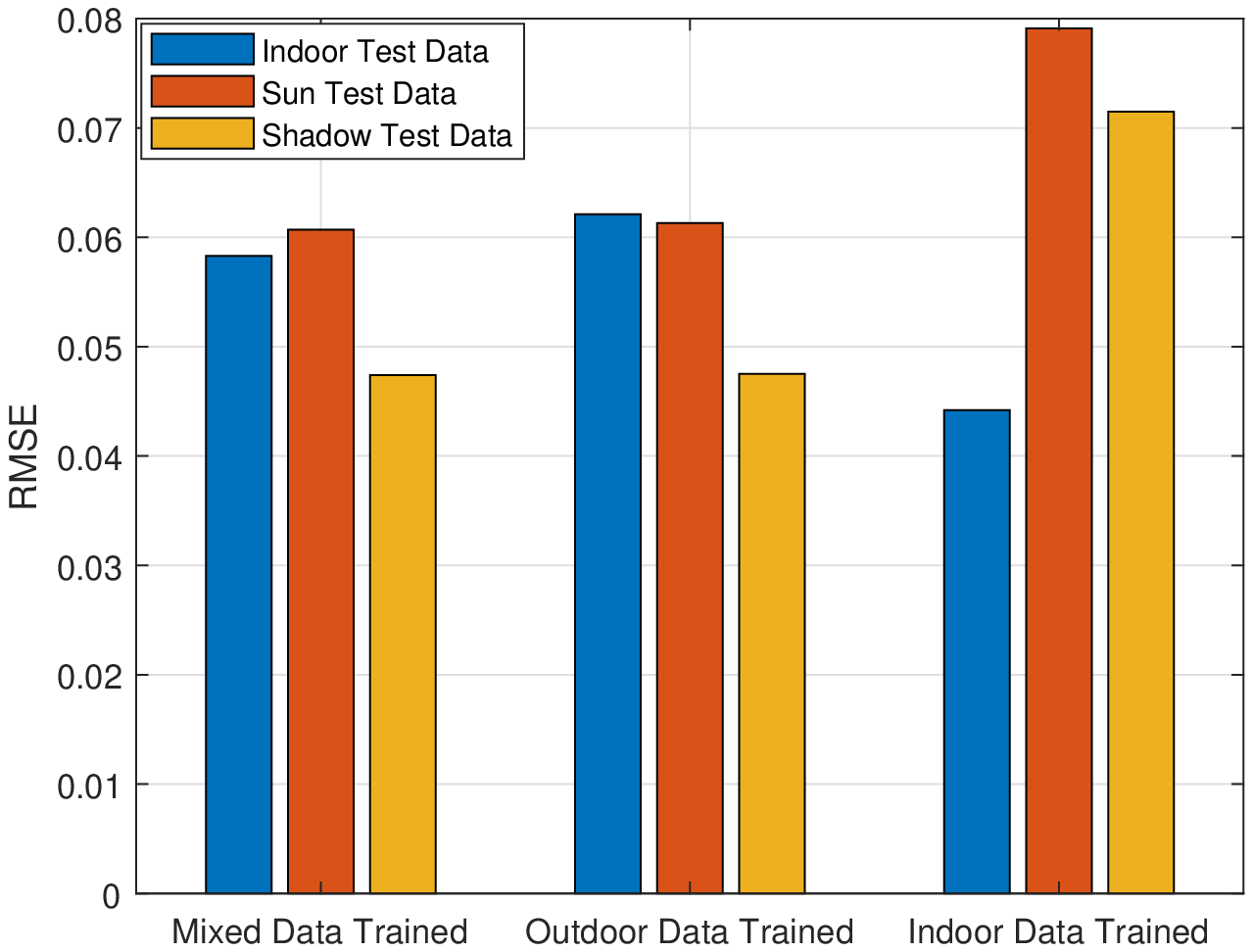}}
    \hfill
  \subfloat[\label{denoiseSamples}]{%
        \includegraphics[clip,trim=0 0 0 20,width=0.5\linewidth]{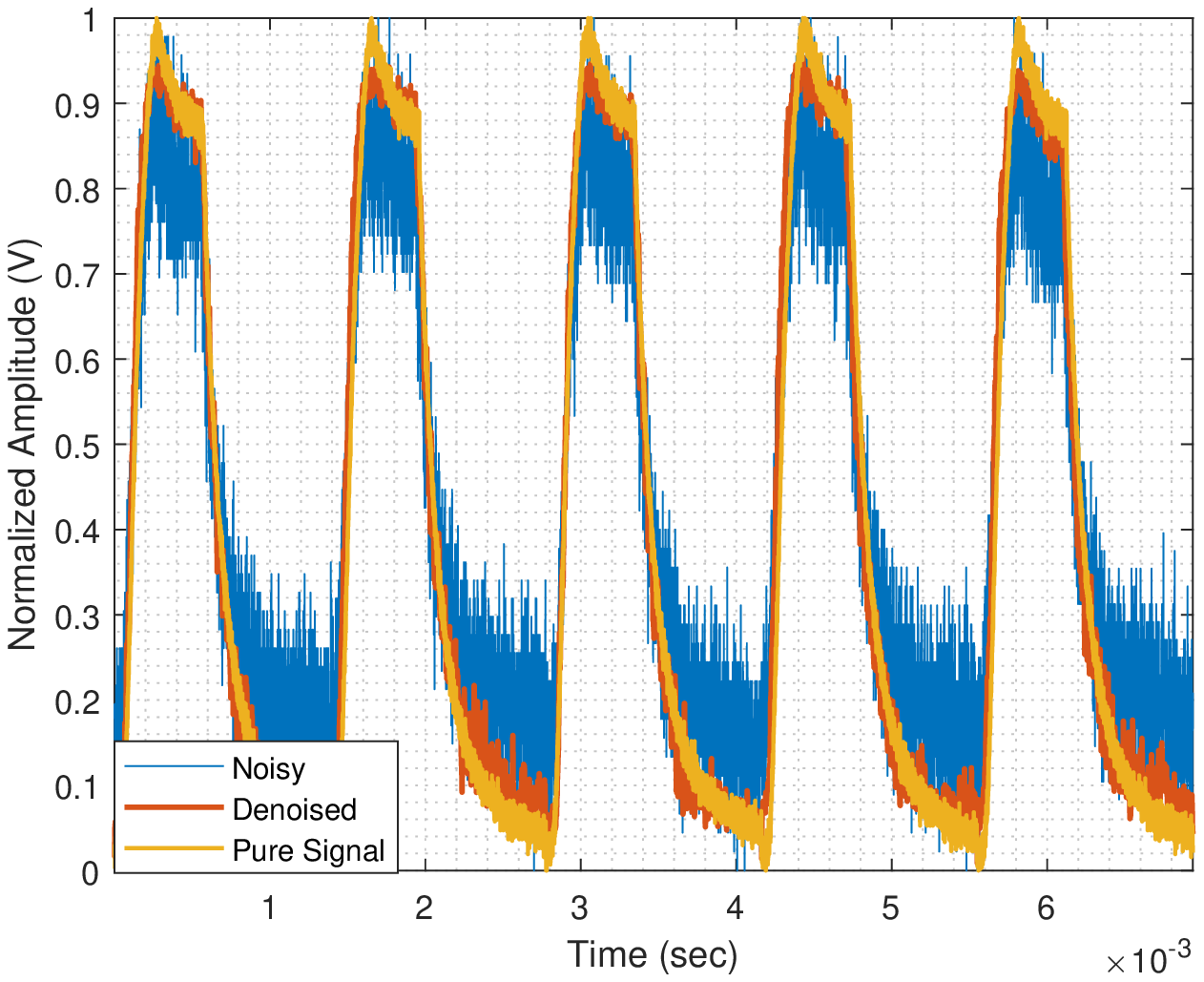}}
       \\
  \caption{(a) Denoising performance of three different data-set trained \glspl{CAE} (b) Noisy, clean and denoised \ac{VVLC} signals for shadow outdoor conditions}
  \label{fig:rmse}
\end{figure}

Fig.~\ref{fig:rmse}~(a) demonstrates the \ac{RMSE} denosing performance of the proposed models for bright sun, shadow outdoor and indoor \ac{LED} light contaminated channels. The proposed \ac{CAE} denoising scheme, trained with MD data-set, achieves 0.0607, 0.0474, 0.0583 \ac{RMSE} for bright sun, shadow outdoor and indoor channels. On the other hand, the model trained with ID data-set yields 0.0791, 0.0715 and 0.0442 \ac{RMSE} for bright sun, shadow outdoor and indoor channels, respectively. Moreover, the OD data-set trained model performs similar to the mixed sample trained model, with 0.0613, 0.0475, 0.0621 \ac{RMSE} for bright sun, shadow outdoor and indoor channels, respectively. Thus, model trained with artificial \ac{LED} background lighting (ID data-set) yields 28.8\% and 22.7\% better denoising performance for \ac{LED} light disturbed symbols when compared to OD trained and MD trained models, respectively. On the other hand, the best performing MD data-set trained model tested with the solar radiation disturbed symbols outperforms the \ac{RMSE} of OD and ID trained models by 0.2\% and 33.7\%, respectively. Therefore, two-different offline trained \glspl{CAE} can be used to denoise \ac{LED} contaminated and outdoor solar radiation contaminated \ac{VVLC} symbols. 

Fig.~\ref{fig:rmse}~(b) presents the outdoor solar radiation contaminated \ac{VVLC} signals for shadow scenario together with the clean and the best performing \ac{CAE} model denoised counterparts. The mean normalized signal amplitudes of noisy, denoised and original signals are, 0.7597 V, 0.8408 V, 0.8483 V for ON symbols and 0.2368 V, 0.1514 V, 0.1429 V, for OFF symbols, respectively. Thus, the \ac{CAE} denoising scheme considerably increases the detectability of the amplitude levels of noisy signals. 

\section{Conclusion} \label{sec:conc}

In this paper, we propose noise analysis and synthesis for \ac{VVLC} together with a \ac{CAE} based denoising scheme. The noise analysis reveals the time correlation properties of the \ac{VVLC} noise due to \ac{LED} background lighting, where the time correlated colored noise is demonstrated to be composed of flicker noise and random walk noise. The random walk noise is the dominant noise component for indoor garage \ac{LED} and vehicle \ac{LED} disturbed \ac{VVLC} channels, where the flicker noise is observed to be stronger for the channels with higher DC component. Since \ac{VVLC} channels are exposed to different noise components, a flexible algorithm to synthesize noise is proposed. The algorithm generates artificial noise data sets in the form of white noise, flicker noise and random walk noise with different strengths. The presented algorithm can be used to simulate \ac{VVLC} system performance under different noise conditions. 
Moreover, we propose a \ac{CAE} based denoising scheme, which reduces the noise due to \ac{LED} lighting and solar radiation for indoor and outdoor environments, respectively. The proposed denoising scheme, trained with indoor \ac{LED} disturbed and outdoor solar radiation disturbed \ac{VVLC} symbols is demonstrated to perform best for outdoor solar radiation disturbed \ac{VVLC} signals under shadow, where the main noise source is in the form of white noise. Therefore, \ac{CAE} based denoising scheme can be used to increase \ac{VVLC} system performance by increasing the symbol detection probability.

\section{Acknowledgement}

This work was supported by CHIST-ERA grant CHIST-ERA-18-SDCDN-001, the Scientific and Technological Council of Turkey 119E350 and Ford Otosan.

\bibliographystyle{ieeetr}
\bibliography{references}

\end{document}